# All-optical tuning of a quantum dot in a coupled cavity system

Ranojoy Bose[1], Tao Cai[1], Glenn S. Solomon[2] and Edo Waks[1,2*]
[1]Department of Electrical Engineering, University of Maryland, College Park, 20742
[2]Joint Quantum Institute, University of Maryland, College Park, 20742
*edowaks@umd.edu

**Abstract:** We demonstrate a method of tuning a semiconductor quantum dot (QD) onto resonance with a cavity mode all-optically. We use a system comprised of two evanescently coupled cavities containing a single QD. One resonance of the coupled cavity system is used to generate a cavity enhanced optical Stark shift, enabling the QD to be resonantly tuned to the other cavity mode. A twenty-seven fold increase in photon emission from the QD is measured when the off-resonant QD is Stark shifted into the cavity mode resonance, which is attributed to radiative enhancement of the QD. A maximum tuning of 0.06 nm is achieved for the QD at an incident power of 88 µW.

A single quantum dot (QD) embedded in a two-dimensional photonic crystal (PC) resonator has emerged as a promising integrated platform for realizing key components in quantum and classical information processing. The resonant interaction of a QD with a microcavity has been shown to achieve the strong coupling regime, observed through a modification of the QD[1] or cavity[2-4] spectrum. This regime has been used to demonstrate a variety of low-power nonlinear optical effects that may be used for efficient single photon generation[5] and all-optical modulation of light[6,7]. The non-resonant interaction of QDs with microcavity modes has also been extensively studied for applications such as QD spectroscopy[8], quantum state readout[9,10], non-resonant photon emission[11], and large optical Stark shifts[12].

The interaction of QDs with the modes of evanescently coupled cavities (photonic molecules)[13] has also been explored in recent literature. For example, such a photonic molecule in a two dimensional (2D) photonic crystal has been used to create a dual wavelength QD laser[14]. In the context of cavity-QD interactions, these systems offer the possibility of multiple distinct resonances that can interact with a QD that may be spatially confined in one of the two cavity modes. As an example, a photonic molecule comprising of two evanescently coupled micropost cavities has been used to generate an ultra-bright source of entangled photons by deterministically coupling both the exciton and biexciton states of a single QD to the two cavity modes[15].

Here, we present another novel application of a 2D photonic molecule. We show that the emission wavelength of a QD can be optically tuned onto resonance with one mode of the molecule by using a laser on resonance with the other mode, leading to enhanced spontaneous emission through the Purcell effect. This unique method of optically tuning a QD is enabled by the nonlinear optical Stark effect[16]. The technique presented here is useful for all-optical modulation of cavity-QD reflectivity[6,7] and may enable fast modulation of QD-based single photon sources.

Measurements are performed on InAs QDs (with QD density 30 $\mu m^{-2}$) embedded in gallium arsenide (GaAs) photonic crystal structures. The designed photonic crystal device (Figure 1a, b) consists of two coupled linear-defect (L3) cavities with single hole tuning of 0.15$a$ (labeled "A"), where $a$ is the lattice constant and is set to 240 nm[14,17]. The hole diameter is set to 140 nm, and the slab thickness is 160 nm. The cavities are vertically spaced by four layers of the photonic crystal and laterally spaced in the direction of the evanescent tail of the local field profile[14]. Finite difference time domain (FDTD) simulations confirm that that this configuration supports two coupled modes in the cavity system that may be classified as anti-symmetric (figure 1a) and symmetric (figure 1b) with theoretical quality factors (Q) of 50000 and 60000 respectively and a predicted mode splitting of 2 nm. Following this design, photonic crystals were fabricated using electron beam lithography, followed by chlorine-based inductively coupled plasma etching and a chemical wet etch to remove a sacrificial layer of AlGaAs below the photonic crystal. A scanning electron microscope (SEM) image of a fabricated device is shown in Figure 1c.

The fabricated sample was mounted in a liquid helium cryostat and optically excited using a titanium:sapphire (Ti:Saph) laser emitting at 780 nm, above the GaAs bandgap, with a spot-size of 1 $\mu$m. The photoluminescence spectrum at 30K, recorded using a confocal microscope setup and a grating spectrometer, reveals two modes with a spectral separation of 3.43 nm (1.2 THz), as shown in Figure 1d. At 30K, the shorter wavelength mode has a resonance at 932.32 nm, while the longer wavelength mode has a resonance at 935.75 nm. The short-wavelength mode, labeled CM1, exhibits a Q of 11,700 ($\kappa/2\pi$=27.9 GHz), and the long-wavelength mode, labeled CM2, has a Q of 10,400 ($\kappa/2\pi$=31.8 GHz). In addition, a single bright emission is observed at 935.45 nm at 30K and attributed to a QD. The QD resonance can be tuned across CM2 by increasing the temperature from 30 K to 44 K (as shown in the surface plot of Figure 1e). As the QD is tuned across the cavity resonance an anticrossing is observed. The QD and cavity become resonant at 37 K, resulting in a vacuum Rabi splitting of 50 pm, corresponding to $g/2\pi$ of 11.86 GHz[18]. Since $4g>\kappa$, the QD-CM2 system is strongly coupled.

The tuning of the emission wavelength of a QD via a cavity-enhanced optical Stark shift has been previously demonstrated in the single cavity case[12]. In a coupled-cavity system, both coupled modes should be able to induce an optical Stark shift on a the QD. In order to demonstrate this effect, we excite the coupled cavity-QD system with a narrowband tunable laser (NewFocus Velocity) in the spectral region of CM1 and CM2. Driving the system with a detuned laser illuminates the QD and both cavity modes through non-resonant energy transfer[19], enabling us to measure their spectrum under strong laser excitation. In order to suppress the background scatter of the uncoupled laser light, cross-polarization spectroscopy is performed[20]. Reflected laser scatter is further reduced by spatially exciting in one cavity region, and collecting at the other cavity. Figure 2a shows the scattering spectrum as the tunable laser is scanned through CM1 resonance at a power of 88 μW recorded prior to the objective lens, using a QD-CM2 detuning of 0.2 nm. As the tunable laser becomes resonant with CM1, enhanced light emission is observed from both the QD and mode CM2. When the laser and CM1 are resonant, the QD emission wavelength shifts towards CM2. The maximum measured shift is 0.06 nm (20 GHz). It should be noted that both the non-resonant energy transfer and the Stark shifts are observed only when the laser wavelength lies within the spectral bandwidth of CM1, indicating that the cavity is resonantly enhancing the field and that the QD is coupled to mode CM1.

At laser wavelengths longer than CM1, the QD emission is suppressed, and increases again as the laser approaches CM2. Figure 2b plots the cavity spectrum as the tunable laser is swept across mode CM2 using a reduced laser power of 34 μW, at QD-CM2 detuning of 0.2 nm. Similar to Fig. 2a, as the laser becomes resonant with the cavity mode a Stark shift is observed. Since the QD is blue detuned from mode CM2, the shift occurs in the opposite direction, with a maximum shift of 0.06 nm attained on resonance. The single photon nature of the emission attributed to the QD is also confirmed using second-order photon correlation measurements ($g^{(2)}(0)=0.2$) under pulsed excitation at the CM2 mode resonance using a 140ps-lifetime laser with a repetition rate of 76 MHz, as shown in Figure 2c.

The power dependence of the measured Stark shift is shown in Figure 3a, which plots the QD emission wavelength as a function of laser power when the QD is blue-detuned from CM2 by 0.12 nm and the laser is fixed on resonance with CM1. As the power is increased from 200 nW to 88 µW the QD emission wavelength (shown by blue diamonds) Stark shifts towards the cavity mode. The emission wavelength of the QD is calculated by fitting the emission spectrum of the QD and cavity to Lorentzians. The solid line represents a linear regression fit to the QD resonant frequency as a function of power. From the slope of the fit the Stark shift rate is determined to be 0.72 pm/µW.

As the QD is tuned into resonance with CM2, we expect an increase in photon counts due to enhanced spontaneous emission from the QD via the Purcell effect [21,22]. In order to estimate the photon emission due to the resonant interaction of the QD with the cavity, we first measure the cavity emission from CM2 by integrating the raw counts within the full-width half-maximum linewidth of the mode, while the laser (88 µW) scans through CM1. The results are shown in Figure 2c. When the laser is resonant with CM1 and the QD is tuned into resonance with CM2, we measure a cavity emission rate of $4.4 \times 10^5$ counts per second for mode CM2. For comparison, we also plot the case where $\delta=-0.2$ nm (corresponding to figure 2b) where the detuning is much larger than the maximum Stark shift such that the QD remains detuned throughout the scan. This scan, recorded at the same power of the tunable laser therefore provides a measure of background emission, which is dominated by inelastic scattering of pump laser photons into mode CM2. By comparing the two sets of data, we estimate a net contribution of $4 \times 10^5$ photons/s directly from the QD into CM2 when the laser is resonant with CM1. Dividing this number by the maximum value of the total integrated emission from the QD in the far-detuned case of $\delta=-0.2$ nm (plotted in the inset of Figure 2c as a function of laser detuning from CM1), we achieve a radiative emission enhancement factor of 27 when the QD is tuned onto resonance with CM2.

In conclusion, we have presented a novel technique to locally tune the interaction between a QD and a cavity mode using the optical Stark effect by using two evanescently coupled cavity modes coupled to a single QD. We have shown that the system sustains high Q and that strong coupling of a single QD can be achieved. A twenty-seven fold increase in photon counts is observed when the QD is tuned into resonance with a cavity mode using the Stark shift. Further improvements in device performance can be attained by considering cavity designs with more efficient input and output coupling to improve both the excitation and collection efficiencies [23]. This technique could ultimately be useful for modulating cavity-QD interactions on ultra-fast timescales for all-optical switching applications.

Acknowledgements: The authors would like to acknowledge support from a DARPA Defense Science Office Grant (No. W31P4Q0910013), the ARO MURI on Hybrid quantum interactions Grant (No. W911NF09104), the physics frontier center at the joint quantum institute, and the ONR applied electromagnetic center N00014-09-1-1190. E. Waks would like to acknowledge support from an NSF CAREER award Grant (No. ECCS-0846494).

Figure 1| (a) Calculated field profile ($\mathbf{E}_y$) for anti-symmetric mode. (b) Calculated field profile for symmetric mode. (c) SEM of fabricated device. Scale bar: 1 µm. (d) Photoluminescence spectrum at 30K showing two cavity modes as well as the QD used in experiments. (e) Temperature scan of PL emission between 30 K and 44 K.

Figure 2| (Inelastic scattering spectrum using narrowband tunable laser at 27K while laser is scanned near (a) CM1 resonance and (b) CM2 resonance, where $\lambda_l$ corresponds to the laser wavelength and scattering intensity is plotted on a logarithmic scale. (c) Second-order correlation measurement performed with the excitation laser resonant with CM2.

Figure 3| (a) QD wavelength (blue diamonds) as a function of excitation power of tunable laser, when the laser is on resonance with CM1. Solid red line shows fit to QD wavelengths. (b) Inelastic scattering spectrum using narrowband tunable laser at 31K while laser is scanned near CM1 resonance. $\log_{10}$ values are shown. (c) Total integrated emission from cavity at $\delta$=-0.06 (red circles) and $\delta$=-0.2 (blue line) as a function of laser detuning from CM1. Inset: Integrated QD emission (cps) as a function of laser detuning from CM1 (at $\delta$=-0.2) using Lorentzian fits to the QD emission.

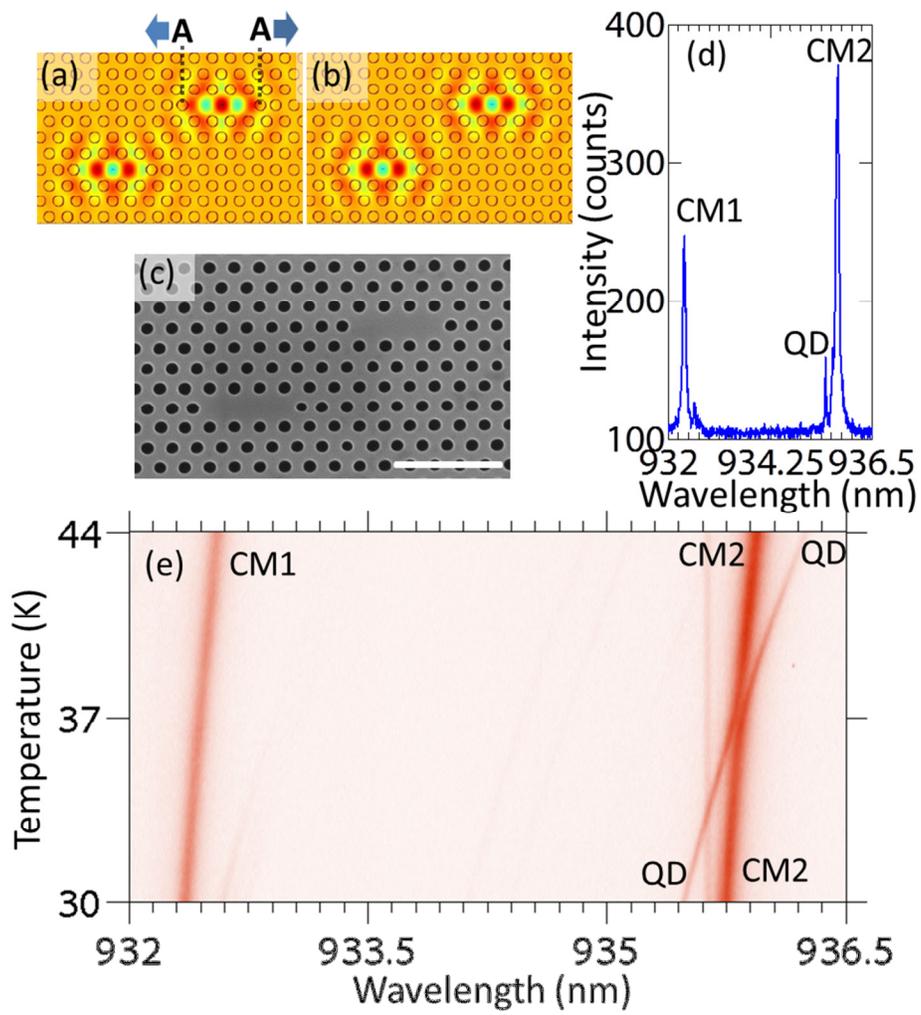

Figure 1

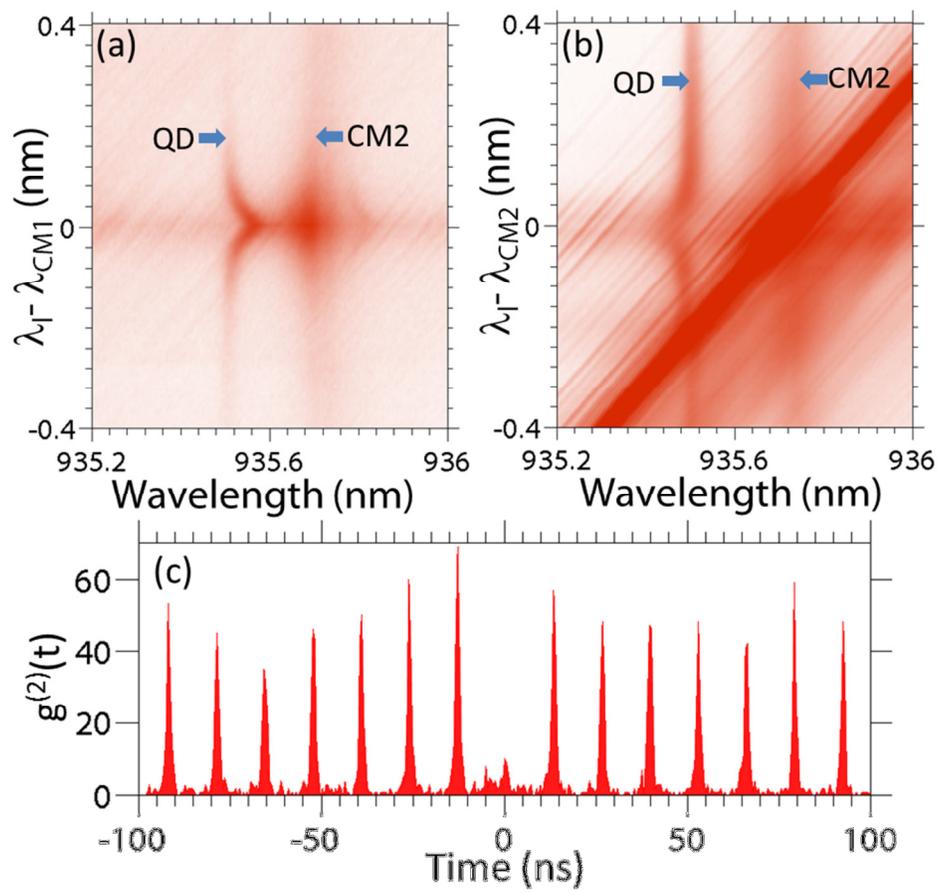

Figure 2

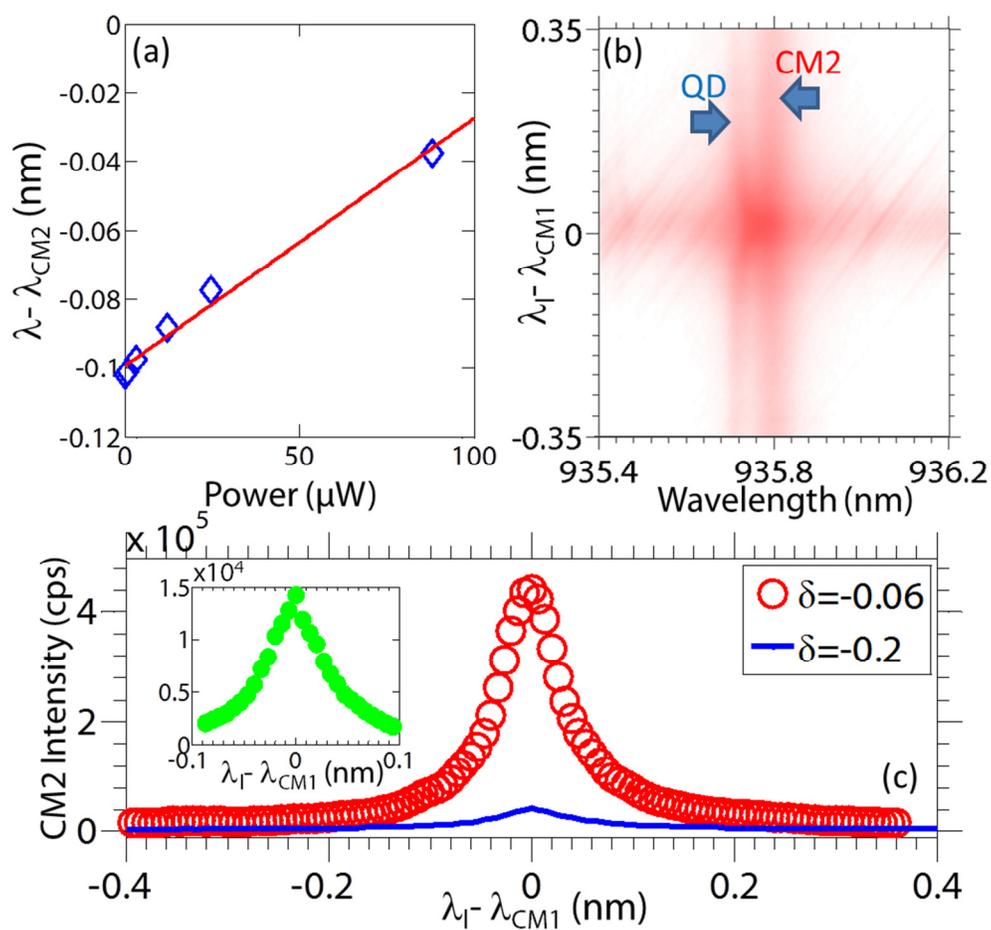

Figure 3